\documentclass{emulateapj}

\slugcomment{}

\shorttitle{Mid-IR spectra of FeLoBALs}
\shortauthors{Farrah et al}

\begin{document}

\title{The extraordinary mid-infrared spectral properties of FeLoBAL Quasars}

\author{D. Farrah\altaffilmark{1}}
\author{T. Urrutia\altaffilmark{2}}
\author{M. Lacy\altaffilmark{3}}
\author{V. Lebouteiller\altaffilmark{4}}
\author{H. W. W. Spoon\altaffilmark{4}}
\author{J. Bernard-Salas\altaffilmark{4}}
\author{N. Connolly\altaffilmark{5}}
\author{J. Afonso\altaffilmark{6,7}}
\author{B. Connolly\altaffilmark{8}}
\author{J. Houck\altaffilmark{4}}

\altaffiltext{1}{Astronomy Centre, University of Sussex, Brighton, UK}
\altaffiltext{2}{Spitzer Science Center, California Institute of Technology, Pasadena, CA 91125, USA}
\altaffiltext{3}{National Radio Astronomy Observatory, Charlottesville, Virginia, USA}
\altaffiltext{4}{Department of Astronomy, Cornell University, Ithaca, NY, USA}
\altaffiltext{5}{Physics Department, Hamilton College, Clinton, NY 13323, USA}
\altaffiltext{6}{Observat\'{o}rio Astron\'{o}mico de Lisboa, Faculdade de Ci\^{e}ncias, Universidade de Lisboa, Tapada da Ajuda, 1349-018 Lisbon, Portugal}
\altaffiltext{7}{Centro de Astronomia e Astrof\'{\i}sica da Universidade de Lisboa, Lisbon, Portugal}
\altaffiltext{8}{Department of Physics and Astronomy, University of Pennsylvania, Philadelphia, PA 19104-6396, USA}

\begin{abstract}
We present mid-infrared spectra of six FeLoBAL QSOs at $1<z<1.8$, taken with the {\it Spitzer} space telescope. The spectra span a range of shapes, from hot dust dominated AGN with silicate emission at 9.7$\mu$m, to moderately obscured starbursts with strong Polycyclic Aromatic Hydrocarbon (PAH) emission. The spectrum of one object, SDSS 1214-0001, shows the most prominent PAHs yet seen in any QSO at any redshift, implying that the starburst dominates the mid-IR emission with an associated star formation rate of order 2700 M$_{\odot}$ yr$^{-1}$. With the caveats that our sample is small and not robustly selected, we combine our mid-IR spectral diagnostics with previous observations to propose that FeLoBAL QSOs are at least largely comprised of systems in which (a) a merger driven starburst is ending, (b) a luminous AGN is in the last stages of burning through its surrounding dust, and (c) which we may be viewing over a restricted line of sight range. 
\end{abstract}


\section{Introduction}
It is now well-established that there exist intimate links between the growth of supermassive black holes and the production of stars in galaxies. Indirect evidence for such links comes from, for example, the tight relationship between central black hole and stellar bulge masses (e.g. \citealt{mag98,geb00,gad09}), and the coeval stellar populations in many massive elliptical galaxies \citep{dun96,ell97,rak07}, suggesting their stars formed within less than a Gyr of each other. Direct evidence comes from the existence of galaxies that simultaneously harbor both high rates of star formation (of order tens to thousands M$_{\odot}$yr$^{-1}$) and rapid accretion onto a central black hole \citep{san96,gen98,far03,lon06}, all enveloped in large quantities of dust, making them extremely luminous in the infrared. The importance of these links for understanding the assembly history of galaxies over at least a substantial fraction of the age of the Universe is demonstrated by the strong evolution of the luminosity function of IR-luminous galaxies with redshift \citep{sau90,lef05}, and the existence of a profusion of IR-luminous sources in the high-redshift Universe (e.g. \citealt{bar98,eal99,cop06,aus09}).

An insightful way to study these links is to identify active galaxies at `key' points, where they are either rapidly building up stellar or central black hole mass, or shifting from one evolutionary phase to the next. One such point may be the "youthful" QSOs; if the starburst precedes or is coeval with the QSO\footnote{The (optical) QSO phase is not necessarily the period in which the SMBH gains the bulk of its mass; indeed, there is evidence that the major periods of BH growth are coeval with or precede the starburst, see e.g. \citet{mar05,mer10,trei10}}, then young QSOs should be those in which the starburst is drawing to a close, and a QSO is starting to emerge from its surrounding dust. Such objects would provide an excellent laboratory for testing feedback mechanisms between the starburst and AGN (e.g. \citealt{sil98,cio07,lag08,moe09,cev09}).

Finding the youthful QSO population is a challenging and ongoing problem (e.g. \citealt{san88,can01,lac06,cop08,geo09,lip09}). Significant attention has focused on the Broad Absorption Line (BAL) QSOs \citep{lyn67,wey91,bro98,sch99,ara01,gre01,hall02,rei03,pri07,gib09}, whose properties\footnote{BAL QSOs come in 3 subtypes. High Ionization BAL QSOs (HiBALs) show absorption in CIV $\lambda$1549, NV $\lambda$1240, SiIV $\lambda$1394 and Ly$\alpha$, and comprise about 10\%\ of all QSOs \citep{tru06}. Low Ionization BAL QSOs (LoBALs) additionally show absorption in MgII $\lambda$2799 and other lower ionization species, and comprise $\sim1.3$\% of all QSOs. Finally, FeLoBAL QSOs, in addition to showing all the absorption lines seen in LoBALs, also show weak iron absorption features \citep{haz87,bec97}. They account for around $0.33\%$ of all QSOs, though the exact fraction is unclear.} have been explained either as arising from being observed at a particular orientation, or because they are young objects still partially surrounded by dust. The properties of HiBALs are now thought to be primarily an orientation effect \citep{surd87,mur95,sch99,gal07,doi09}, while the LoBALs remain controversial, with evidence favoring both orientation and evolution (e.g. \citealt{voi93,ogl99,gal07,gho07,mon08,urr09,zha10}). For FeLoBALs there is also controversy, but the evidence favoring the evolution scenario is (arguably) stronger than for the LoBALs. Based on rest-frame UV and optical spectra, Hall et al (2002) suggest that FeLoBALs are young objects still enveloped in dust. Similar conclusions are reached by Gregg et al (2002), who also postulate that FeLoBALs may be associated with mergers. Further evidence comes from the discovery of iron absorption lines in the UV spectra of two low-redshift ULIRGs which harbor obscured AGN \citep{far05}. Finally, mid-IR photometry observations (\citealt{far07a}, hereafter F07) found that many FeLoBAL QSOs were IR-luminous, and may harbor high rates of star formation. 

Excellent distinguishing evidence of the `youth vs. orientation' debate for FeLoBAL QSOs would be observations that show directly that high rates of star formation accompany the AGN. Given their reddened nature, a good place to look for such evidence is in the md-infrared, in which the spectral shapes of reddened AGN differ markedly from those of star-forming regions. The outstanding capabilities of the Infrared Spectrograph (IRS, \citealt{hou04}) on-board the Spitzer space telescope \citep{wer04} provided a huge step forward in available mid-IR spectroscopic capabilities, with the opportunity to shed light on the FeLoBAL phenomenon. In this paper, we use Spitzer-IRS observations of six FeLoBAL QSOs to examine the idea that they are young objects. We assume a spatially flat cosmology, with $H_{0}=70$ km s$^{-1}$ Mpc$^{-1}$, $\Omega=1$, and $\Omega_{\Lambda}=0.7$.

\section{Sample Selection}
We selected our sample in early 2006, with the requirements that candidates be confirmed as FeLoBAL QSOs via rest-frame UV spectroscopy, and lie in a redshift range where we can observe useful diagnostics with the IRS. Accordingly, we chose six objects (Table \ref{tabsample}) at random from F07. The F07 sample is drawn randomly from the FeLoBAL QSO population known at the time, lie in the redshift range $1.0<z<1.8$, thus placing important mid-IR spectral features in the IRS bandpass, and have photometry observations at 24$\mu$m, 70$\mu$m and 160$\mu$m, which provide good constraints on IR luminosities. 

A downside of this selection though is that our sample is heterogeneous. When we selected our sample, the few known FeLoBAL QSOs had been found in several different surveys. Since then, larger, homogeneous samples of FeLoBAL QSOs have been published \citep{tru06,sca09}, but these samples were not available to us. Accordingly, some level of bias is inevitable in our sample, though it is difficult to quantify what effect this may have. We therefore simply list the origins of our sample; three objects were found via the Sloan Digital Sky Survey \citep{hall02,ade08}, one (ISO 0056-2738) was discovered serendipitously from followup of distant clusters \citep{duc02}, and two (SDSS 1427+2709 \& SDSS 1556+3517) were discovered during spectroscopic followup of quasars from the FIRST survey \citep{bec97,naj00}. SDSS 1556+3517 is radio-loud, while the other five are radio quiet. 

Furthermore, at least three of our sample appear to have atypical\footnote{We include the qualifier as the rest-frame UV absorption line properties of FeLoBAL QSOs have not been studied exhaustively} iron absorption features \citep{hall02}. SDSS 1154+0300 has troughs where the absorption remains significant at velocities comparable to the spacing between absorption features (so $\gtrsim$12,000 km s$^{-1}$), causing them to overlap each other. SDSS 2215-0045 and possibly SDSS 1214-0001 have stronger Fe{\sc III} absorption than Fe{\sc II} absorption, suggesting that the gas in which the BALs occur is dense, hot, and moderately highly ionized. We therefore do not correlate mid-IR spectral properties with those of the iron absorption features.

\section{Methods}
We observed with the IRS using the first order of the short-low module, and both orders of the long-low module, giving observed-frame wavelength coverage of 7.5$\mu$m to 35$\mu$m. As the 24$\mu$m fluxes of our sample span a range of values, we used different observation times for each object so as to to give a signal-to-noise of at least 15 in the continuum at observed-frame 24$\mu$m. Observations were performed in staring mode, using `high' accuracy peak up observations performed with the blue array from a nearby Two Micron All-Sky Survey (2MASS, \citealt{skr06}) star. 

The data were processed through the {\it Spitzer} Science Center's pipeline software (version 18.7), which performs standard tasks such as ramp fitting and dark current subtraction, and produces Basic Calibrated Data (BCD) frames. Starting with these frames, we produced reduced spectra using the SMART v8.0 software package, following the methods described in \citet{leb09}, which we summarize here. Individual frames were cleaned of `bad' pixels using the IRSCLEAN task. The first and last five pixels, corresponding to regions of reduced sensitivity on the detector, were then removed. The individual frames at each nod position were then median combined with equal weighting on each resolution element. Sky background was removed from each image by subtracting the image for the same order taken with the other nod position (i.e. `nod-nod' sky subtraction). One-dimensional spectra were then extracted using `optimal' extraction with default parameters, and defringed using the internal SMART algorithm. We found that, in all cases, the sources were point-like, with a FWHM that was never wider than the PSF. We also checked that the sources were centered in the IRS slit by extracting spectra using `simple' (i.e. not PSF weighted) extraction. We found that the resulting spectra were of slightly lower signal-to-noise but consistent with the optimally extracted spectra, confirming that any slit offset is insignificant.

This procedure results in separate spectra for each nod and for each order. The spectra for each nod were inspected; features present in only one nod were treated as artifacts and removed. The two nod positions were then combined. As the two nod positions have slightly offset wavelength grids, we combined the nods by first interleaving them, and then interpolating the fully sampled spectrum onto a reference wavelength grid. The nod-combined spectra in the three orders were then merged to give the final spectrum for each object. Overall, we obtained excellent continuum matches between different orders. Only in one case (SDSS 1556+3517) was there a significant order mismatch, between the SL1 and LL2 orders. For this object, we scaled the SL1 spectrum by a factor of 1.10 to match the blue end of the LL2 continuum

There remain however several uncertainties over error propagation in the IRS reduction process. For example, defringing is still not completely understood, so some residuals likely remain that are not included in the `formal' errors. Another example is the order mismatch between SL1 and LL2 for SDSS 1556+3517 - a PSF weighted 'optimal' extraction should in principle not produce this effect, and while the scaling is small, we do not  understand why it is necessary. Overall therefore, we regard the resulting errors on each resolution element to be somewhat smaller than they should be, though the degree of this underestimate is likely insignificant.

\section{Results}
\begin{figure}
\begin{minipage}{180mm}
\includegraphics[angle=90,width=85mm]{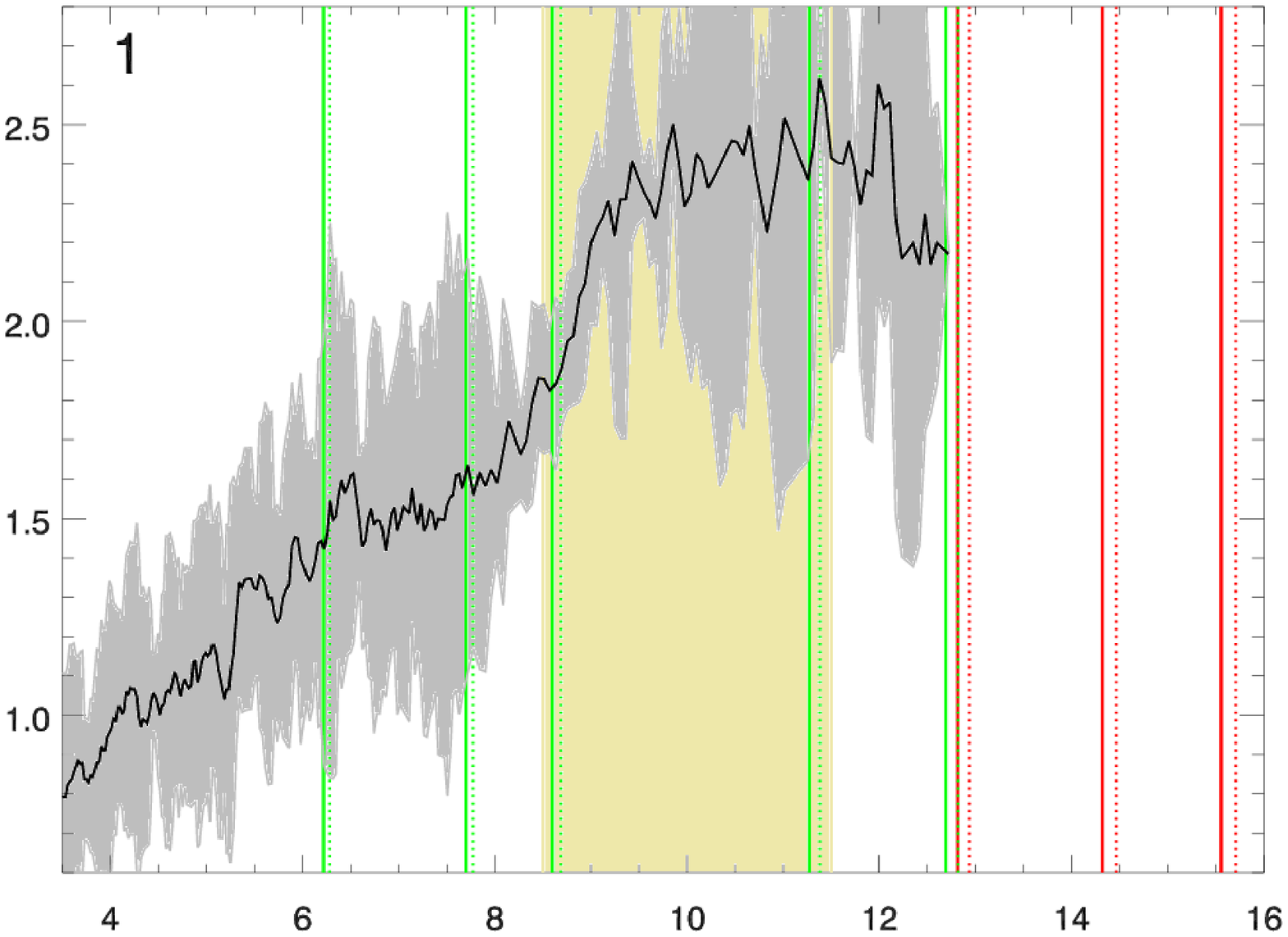}
\includegraphics[angle=90,width=85mm]{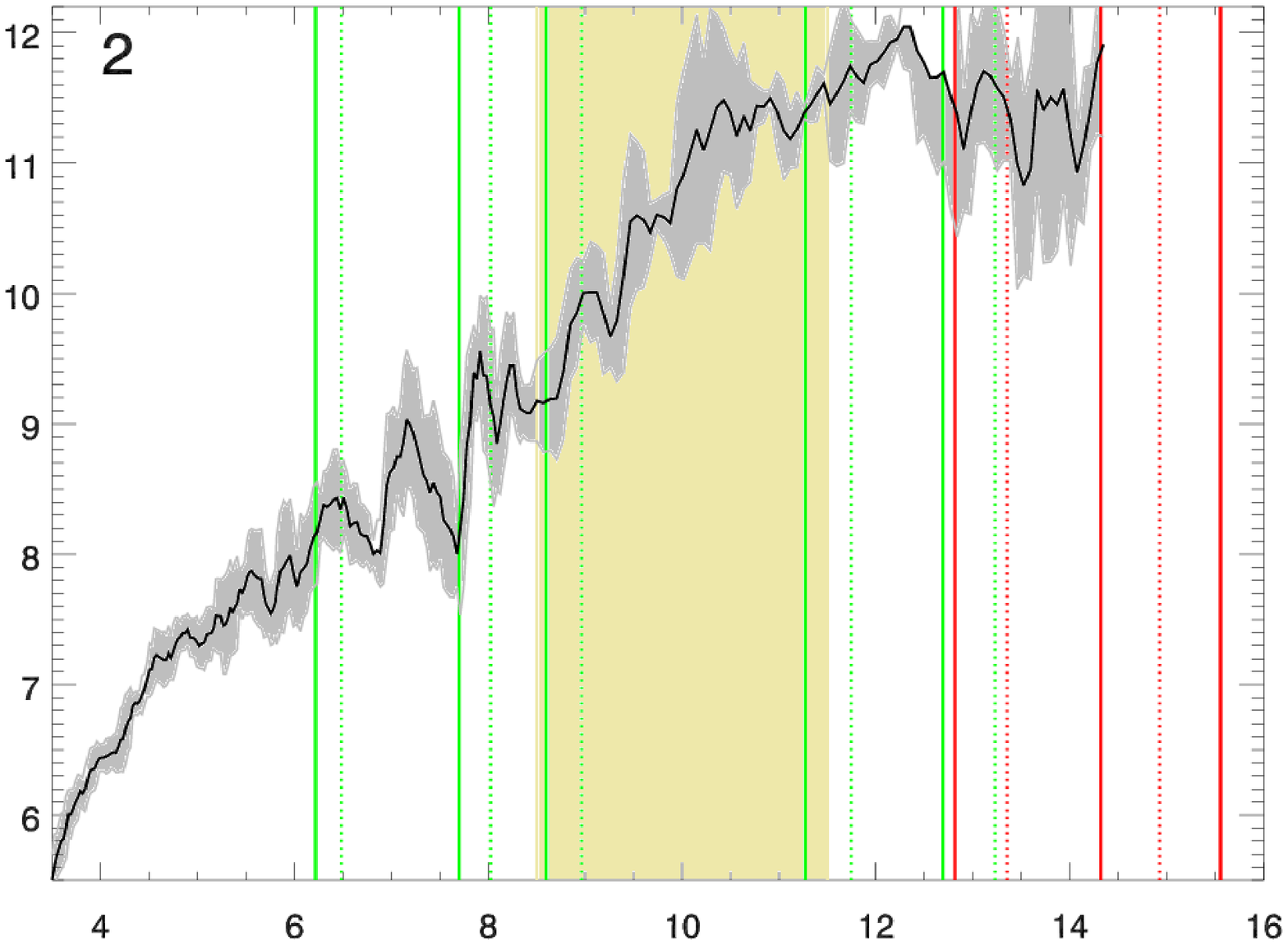}
\end{minipage}
\begin{minipage}{180mm}
\includegraphics[angle=90,width=85mm]{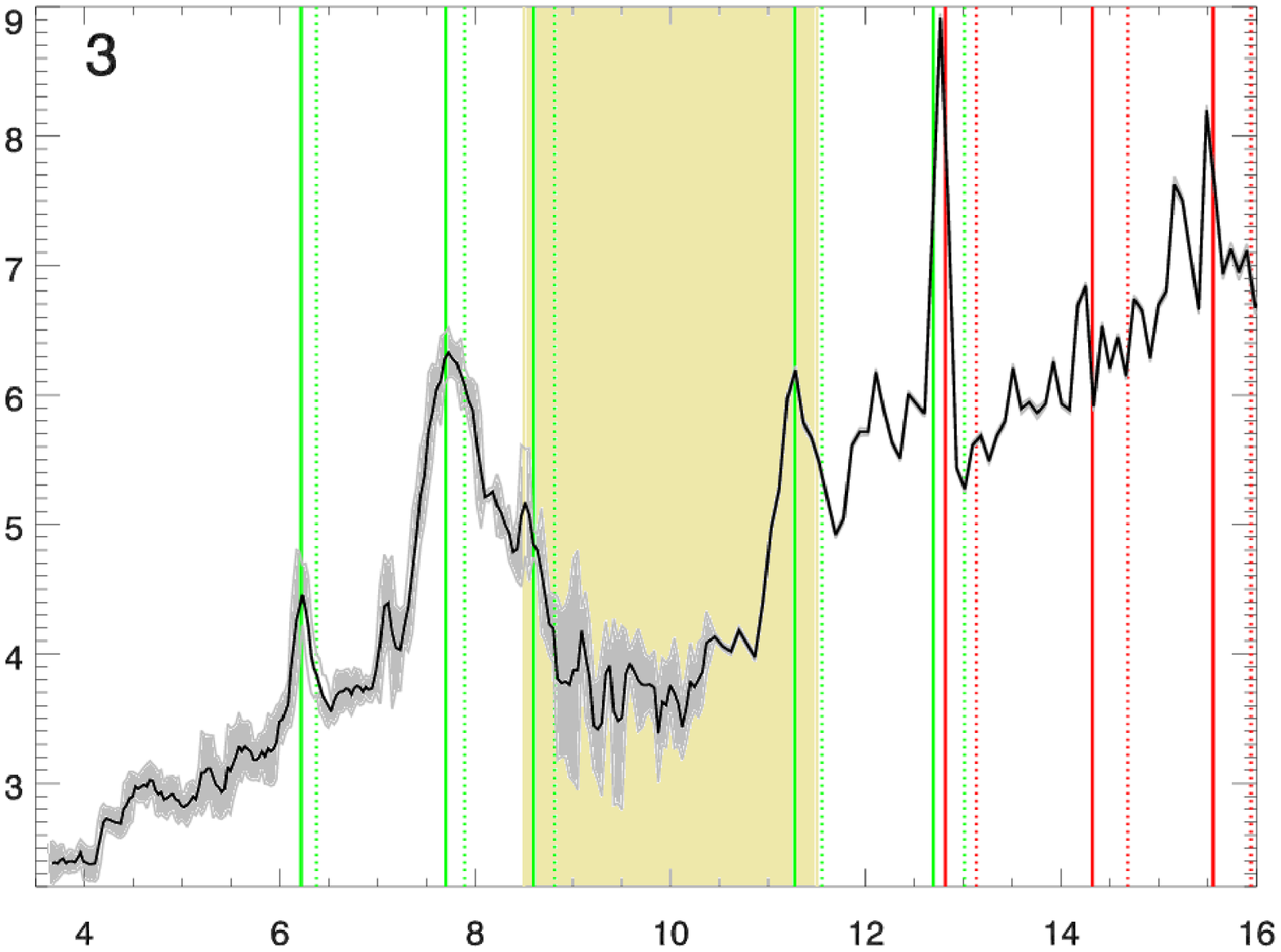}
\includegraphics[angle=90,width=85mm]{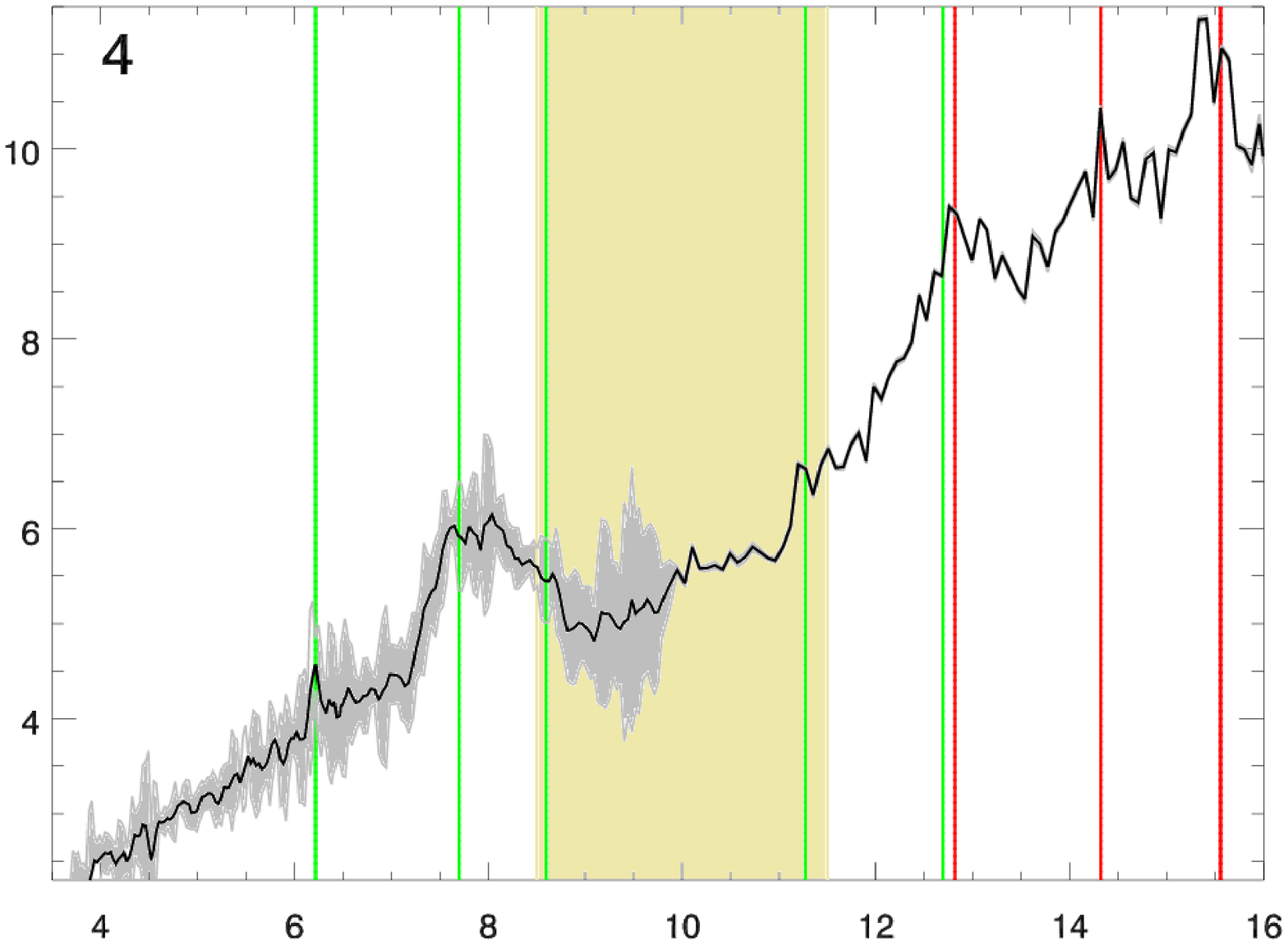}
\end{minipage}
\begin{minipage}{180mm}
\includegraphics[angle=90,width=85mm]{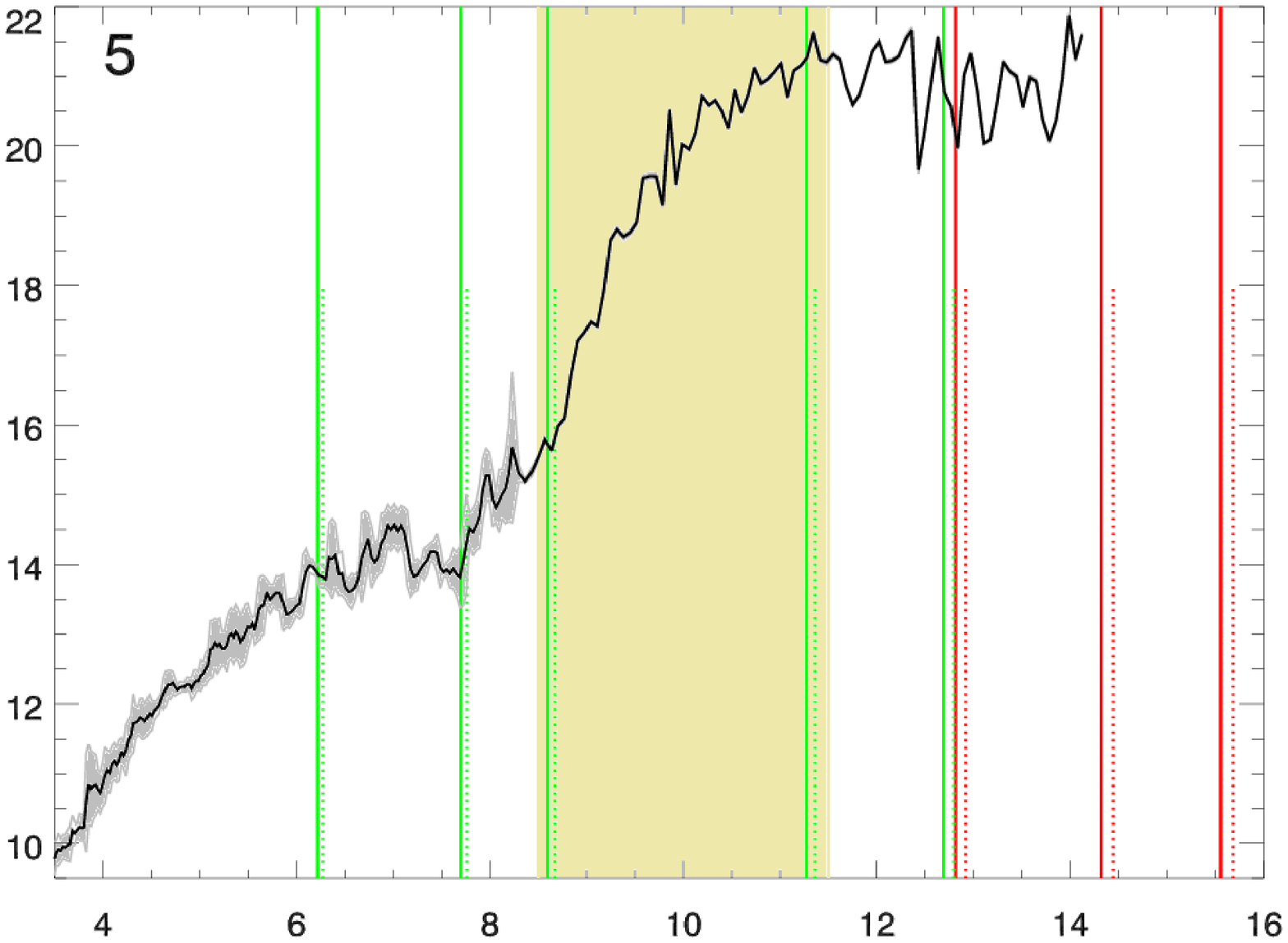}
\includegraphics[angle=90,width=85mm]{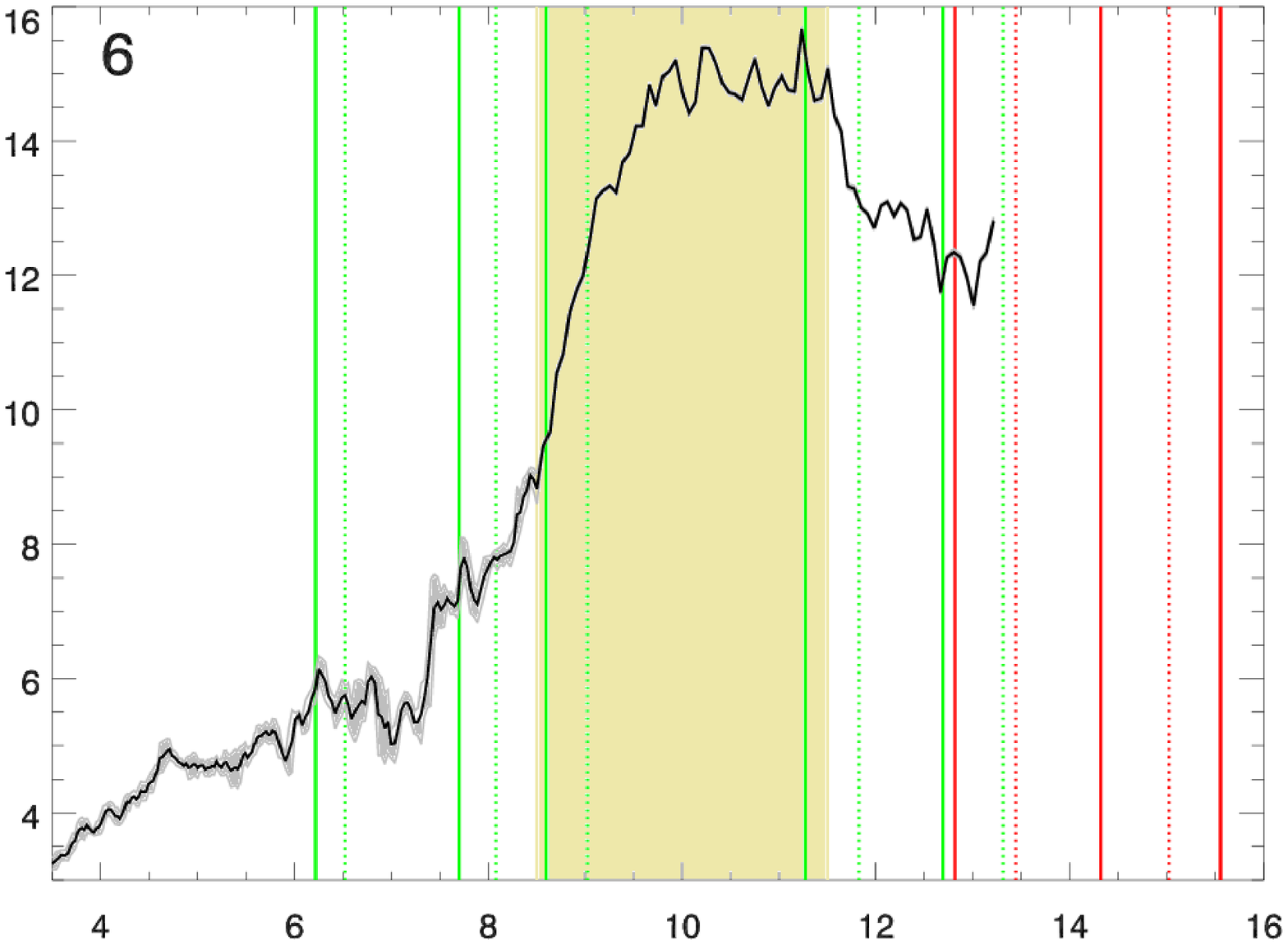}
\end{minipage}
\caption{Spitzer-IRS low-resolution spectra of our sample, plotted in the rest-frame using the optical emission line redshifts. The data are plotted in black, with 1$\sigma$ errors in grey. Flux densities are in mJy and wavelengths are in $\mu$m. The number in the top left of each panel is the ID number in Table \ref{tabsample}. The green solid (dashed) lines mark the 6.2$\mu$m, 7.7$\mu$m, 8.6$\mu$m, 11.2$\mu$m and 12.7$\mu$m PAH features assuming the systemic (peak absorption, if available) redshift. The red lines perform the same function for the [Ne II]$\lambda$12.81, [Ne V]$\lambda$14.32 and [Ne III]$\lambda$15.56 lines. The yellow shaded region shows the approximate extent of the 9.7$\mu$m silicate absorption feature, but see \citealt{bow98,dra03,nik09}. 
\label{spectra}}
\end{figure}

The IRS spectra are presented in Figure \ref{spectra}. Spectral measurements are presented in Table \ref{tabpahsils}. We found one further object serendipitously in the slit of ISO 0056-2738, which appears to be a PAH dominated system at $z\simeq1.42$, but do not consider it further in this paper.

\subsection{Spectral Features}
\begin{figure}
\begin{minipage}{180mm}
\includegraphics[angle=90,width=130mm]{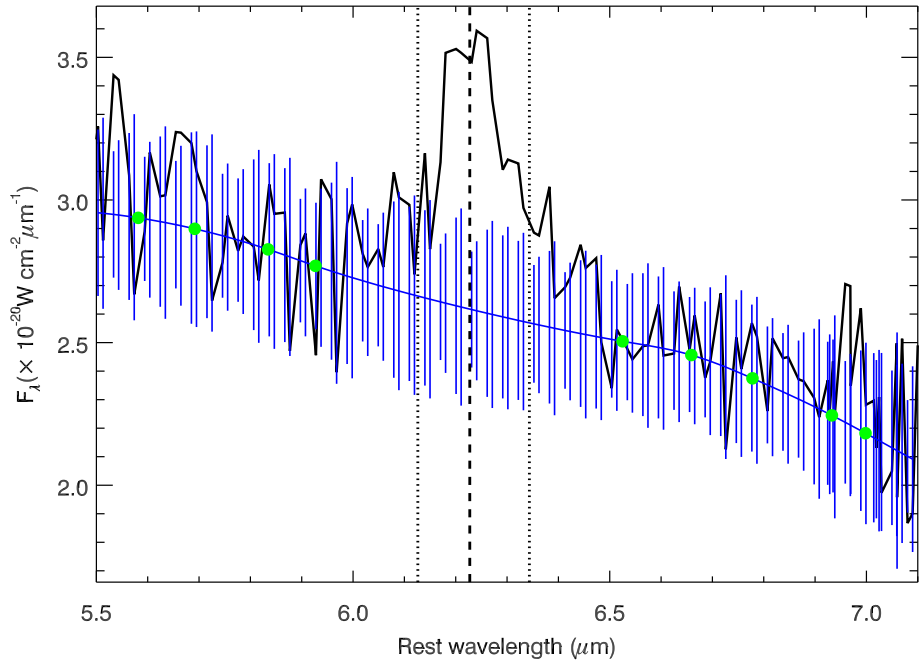}
\end{minipage}
\caption{An example of the spline fits used to determine the shape of the underlying continuum for the PAH fluxes, in this case for the 6.2$\mu$m PAH feature in SDSS 1214-0001. The (rest-frame) spectrum is plotted in black. The points used to define the continuum are shown in green, while the spline fit and its residuals are shown in blue. The errors on the spectrum have been omitted for clarity. The dashed line shows the adopted peak wavelength of the PAH feature, while the dotted lines show the lower and upper limits of the integration. This plot also demonstrates the excellent agreement between the two nods; the `double' blue error bars show separately the error bars for the two nod positions, which have slightly offset wavelength grids. 
\label{examplefit}}
\end{figure}

The spectra show a variety of spectral features. Four objects show one or more broad emission features at 6.2$\mu$m, 7.7$\mu$m, 11.2$\mu$m and 12.7$\mu$m, attributed to bending and stretching modes in Polycyclic Aromatic Hydrocarbons (PAHs, the 12.7$\mu$m feature also contains a contribution from [NeII]$\lambda$12.81). The redshifts determined from the PAHs were in all cases consistent with the optical emission line redshifts, rather than the redshifts at which the UV absorption features peak, placing the source of the PAH emission in the host galaxy rather than in the outflow. The PAH fluxes and equivalent widths (EWs) were computed by integrating the flux after subtracting off a spline interpolated local continuum. An example of the spline fits is shown in Figure \ref{examplefit}.

At least three objects contain a broad feature centered approximately at 9.7$\mu$m, seen in both emission and absorption, that arises from an Si-O stretching mode in Silicate dust \citep{kna73}. We measured the strengths of these features via:

\begin{equation}
S_{sil} = ln\left(  \frac{F_{obs}(9.7\mu m)}{F_{cont}(9.7\mu m)} \right)\label{silstrength}
\end{equation}

\noindent where $F_{obs}$ is the observed flux density at rest-frame 9.7$\mu$m, and $F_{cont}$ is the flux at the same wavelength deduced from a spline fit to the continuum on either side \citep{spo07,lev07,sir08}.

We also searched for other features seen in IR-luminous systems, though these are reliably measurable only with higher resolution data, so we do not present flux measurements. Both SDSS 1214-0001 and SDSS 1427+2709 show strong [NeIII]$\lambda$15.56, and weak but significant [NeV]$\lambda$14.32. SDSS 1214-0001 additionally has weak detections at the positions of [ArII]$\lambda$6.99 and H$_{2}$S(2)$\lambda$12.29. The other four objects show no further features that we can identify. SDSS 1154+0300 has an apparent feature at rest-frame $\sim8\mu$m that we cannot find a reliable ID for, though as the spectrum is low S/N it is possible this feature arises from the juxtaposition of a declining PAH and a rising silicate feature. ISO 0056-2738 is too low S/N to infer the presence or otherwise of other features. We also note that all six spectra likely contain residual structure due to fringing that the defringing algorithm was unable to completely remove.

\subsection{Qualitative Comparisons}

\subsubsection{Objects 3 \& 4}
\begin{figure}
\begin{minipage}{180mm}
\includegraphics[angle=90,width=85mm]{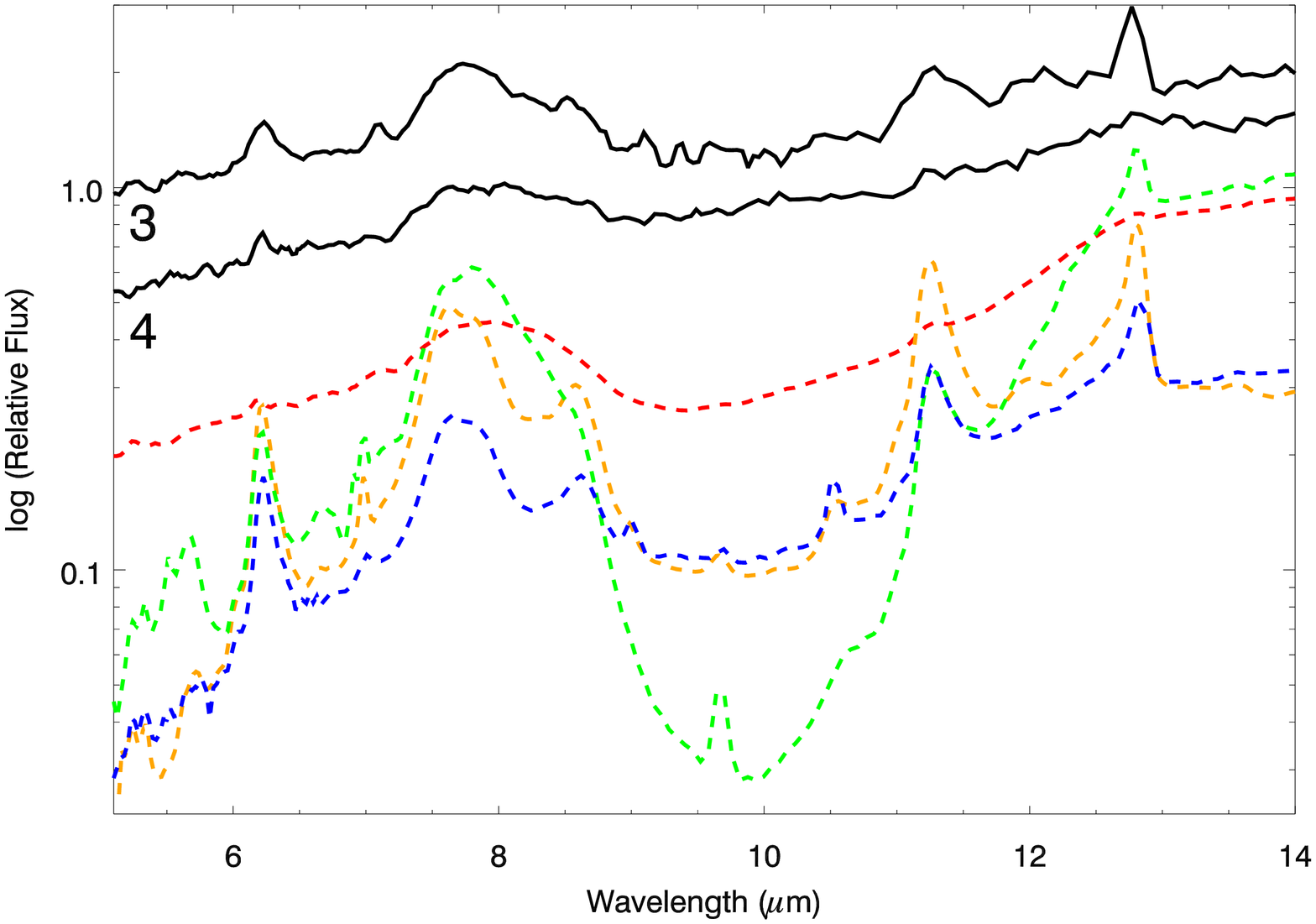}
\includegraphics[angle=90,width=85mm]{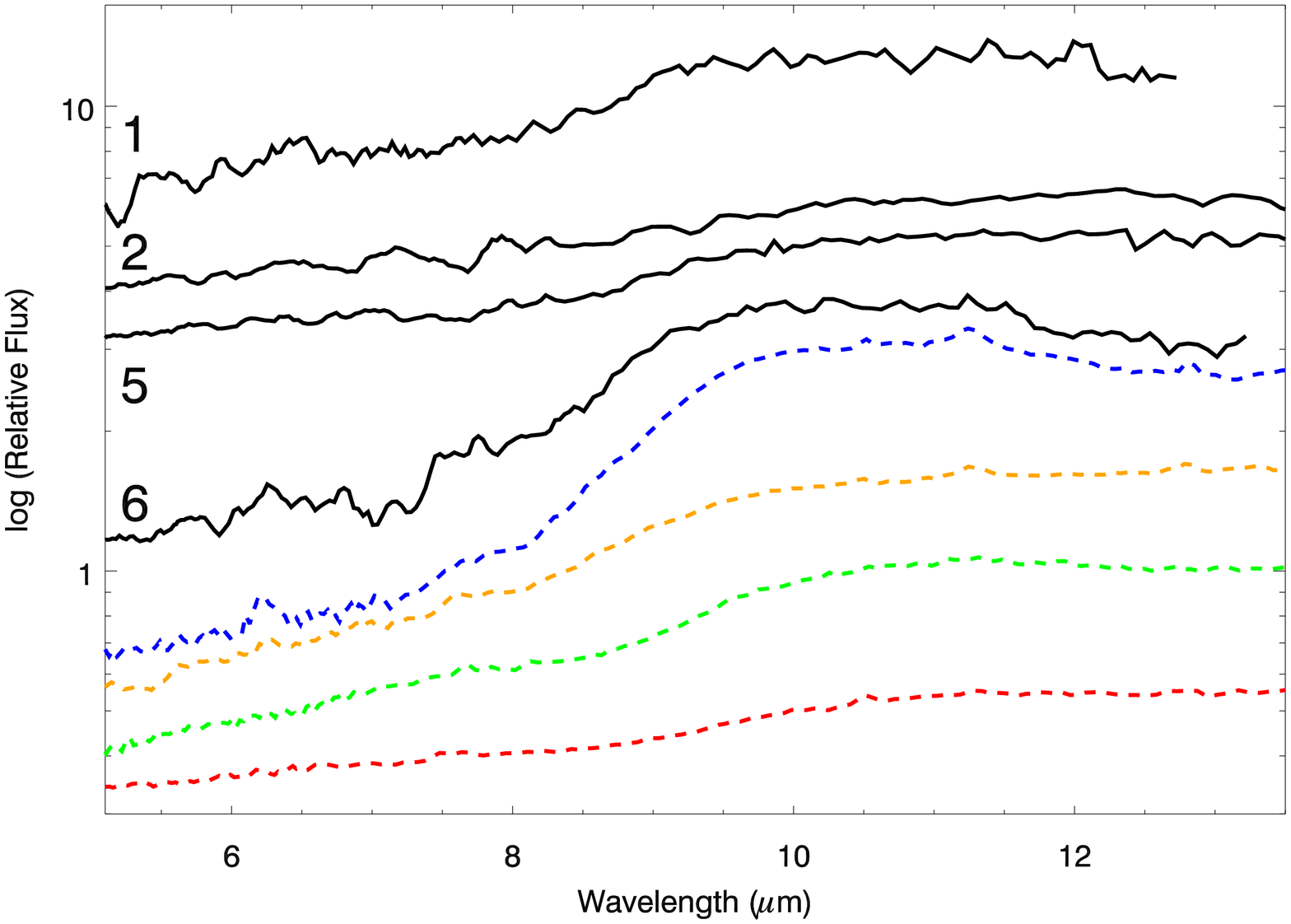}
\end{minipage}
\caption{Comparison of our spectra to well studied low-redshift objects. Objects showing silicate absorption in the left panel, and the others are in the right panel. Comparison spectra are taken from \citet{bdl06,arm07,shi07} and \citet{ima07}. {\it Left:} Red - Mrk 231. Orange - composite of local $\simeq10^{11}$L$_{\odot}$ starbursts. Green - Arp 220. Blue - IRAS 15206+3342 (a local ULIRG with iron absorption features in its UV spectrum). {\it Right:} Red - 3C 273. Orange - I Zw 1 (a low-z NLS1). Green - PG 1211+143. Blue - PG 1351+640 (both far-IR luminous QSOs).
\label{comparisons}}
\end{figure}

We start by comparing the two objects with prominent PAHs, SDSS 1214-0001 and SDSS 1427+2709, to well studied low-redshift objects (Figure \ref{comparisons}). Both objects closely resemble PAH dominated ULIRGs. It is superficially interesting that SDSS 1214-0001 is a good match to IRAS 15206+3342, a local ULIRG with iron absorption features in its UV spectrum, though not of great significance given that many local ULIRGs have similar mid-IR spectral shapes, see e.g. the position of IRAS 15206+3342 in the `network' plot of \citet{far09b}. SDSS 1427+2709 on the other hand is well matched by ULIRGs with weaker PAHs and a more pronounced continuum, such as Mrk 231. Neither system resembles heavily absorbed ULIRG spectra such as Arp 220. 

Moving on to comparisons with larger samples, we are hampered as there does not exist a comprehensive mid-IR spectroscopic survey of HiBALs or LoBALs to compare to. Even comparing to the general AGN population though, the peculiarity of these two objects, in particular SDSS 1214-0001, is thrown into sharp relief. Their PAHs are extraordinarily prominent, far more so than for any QSO, radio loud or radio quiet, that we are aware of \citep{haa05,hao05,shi06,mai07}, including those selected to be far-IR luminous \citep{lut08,mar08}, as well as the X-ray luminous `type 2 QSO' objects \citep{stu06}. There are however a few systems with prominent PAHs among the Narrow Line Seyfert 1 (NLS1, objects with H$\beta$ FWHMs of $<2000$km s$^{-1}$, \citealt{ost85}) population \citep{san10}, and several examples of Sy2 ULIRGs with strong PAHs (\citealt{arm07,ima07,far07b}, see also the type 2 object LH901A in \citealt{stu06}). 

\subsubsection{Objects 1, 2, 5 \& 6}
Turning to the other four sources; SDSS 2215-0045 is a good match to QSOs with strong silicate emission and weak but detectable PAHs, of which several exist \citep{hao05}. Interestingly, SDSS 2215-0045 is an excellent match to PG1351+640, a IR-luminous, CO detected QSO with narrow BALs, FeII in emission under H$\beta$, and slight morphological disturbance \citep{gel94,fal95,zhe01,eva01}. Conversely, SDSS 1556+3517, with its weaker silicate emission feature and negligible PAHs, is a good match both to some classical QSOs, and to NLS1 systems such as PG1211+143 (which has a high velocity outflow, \citealt{pou06,shi07}) and IZw1 (which has optical and UV FeII {\it emission} and a decaying starburst \citep{mar96,sch98,sur98}). SDSS 1154+0300 has a poorer quality IRS spectrum but is also consistent with QSOs with weak silicate emission and negligible PAHs. ISO 0056-2738 is of  too low signal-to-noise to draw any conclusions other than it appears to be consistent with AGN generally. 

\subsubsection{The sample as a whole}
Finally, we compare the ensemble properties of our sample to other classes of active galaxies. We start by looking for matches with IR-selected AGN samples. One sample where we might expect overlap is with the X-ray detected LIRGs in \citet{bra08}, but the two samples do not resemble each other. None of the 16 \citealt{bra08} objects have detectable PAHs; though 9 have some silicate absorption, the rest are featureless power laws. Our two PAH dominated sources resemble some sources in optically faint 70$\mu$m selected samples \citep{bra08a,far09a}, but 70$\mu$m selected samples have no sources with silicates in emission. We do better if we instead compare to IRS observations of high-redshift 15$\mu$m selected samples \citep{her09}, likely because we're selecting on hotter dust; 15$\mu$m samples contain objects comparable to our PAH strong objects, but have few sources with silicates in emission. The one optically selected sample that appears reasonably matched to ours is the NLS1 sample of \citet{san10}, which contains both PAH dominated objects, and a few objects with what appears to be weak silicate emission.

\subsection{Spectral Diagnostics}\label{specfeat}
We move on to quantitative diagnostics. As our spectra are of relatively low S/N and do not have coverage in all the IRS modules, we use simple diagnostics that allow for easy comparisons with other samples. 

\begin{figure}
\begin{minipage}{180mm}
\includegraphics[angle=90,width=85mm]{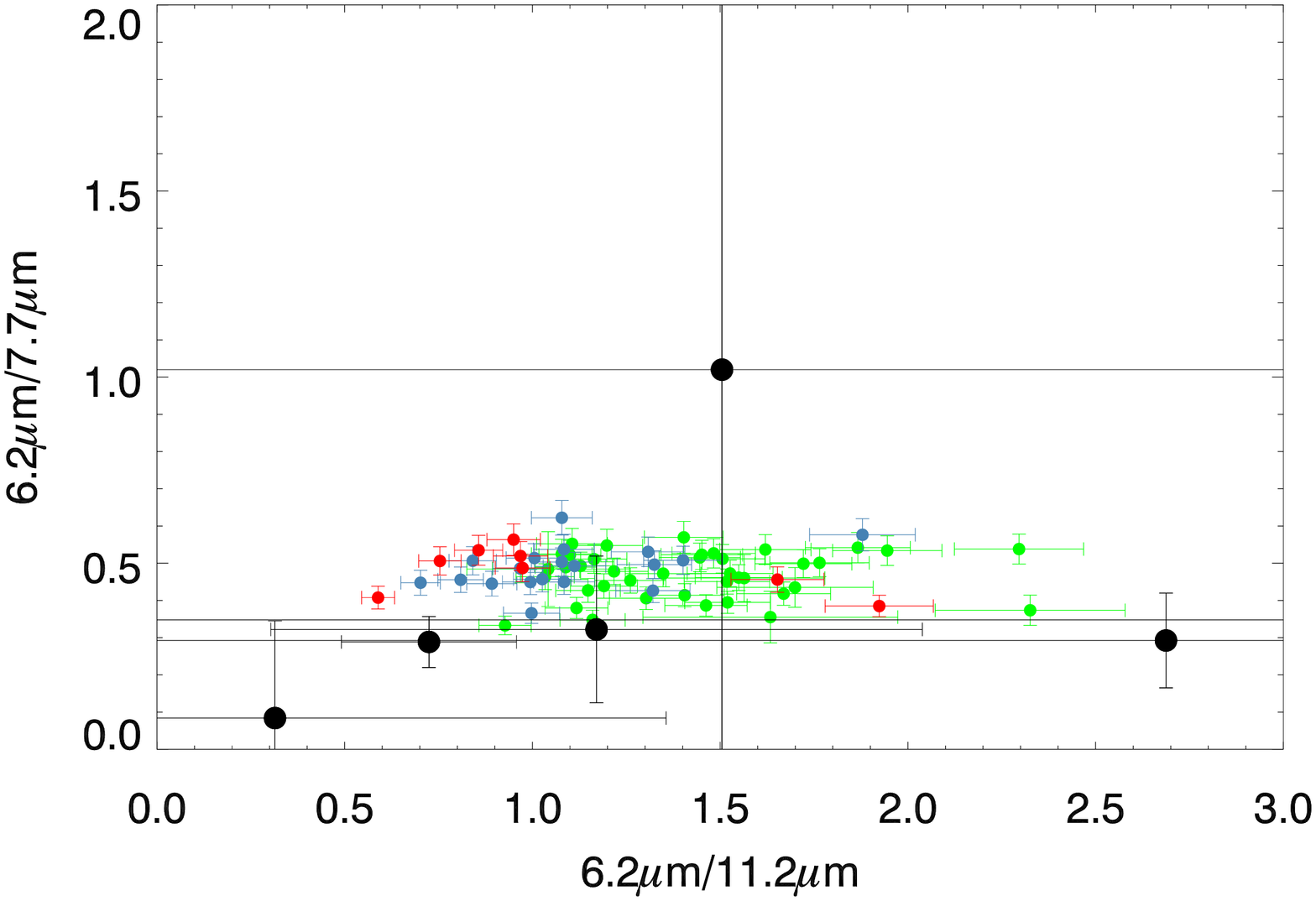}
\includegraphics[angle=90,width=85mm]{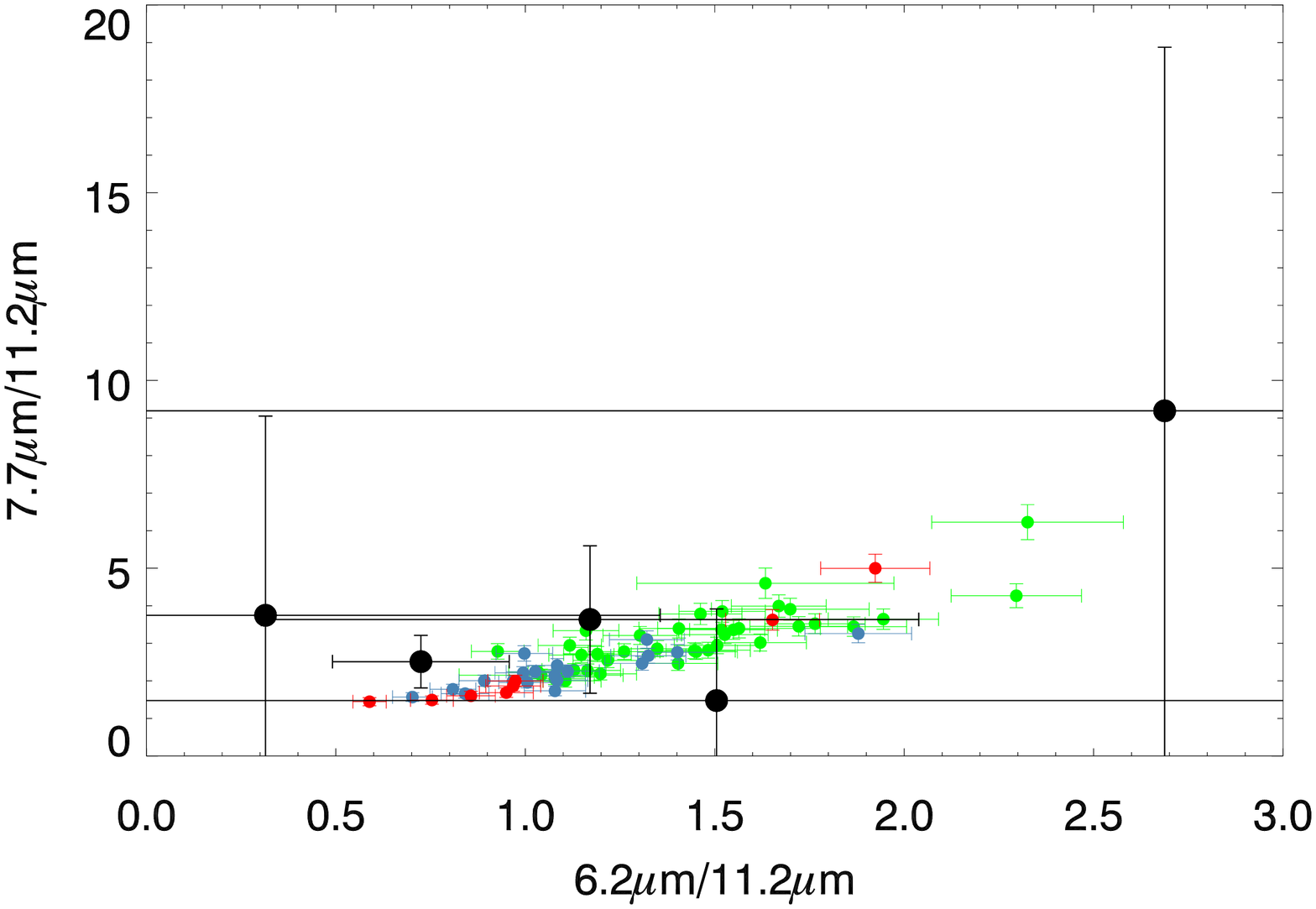}
\end{minipage}
\begin{minipage}{180mm}
\includegraphics[angle=90,width=85mm]{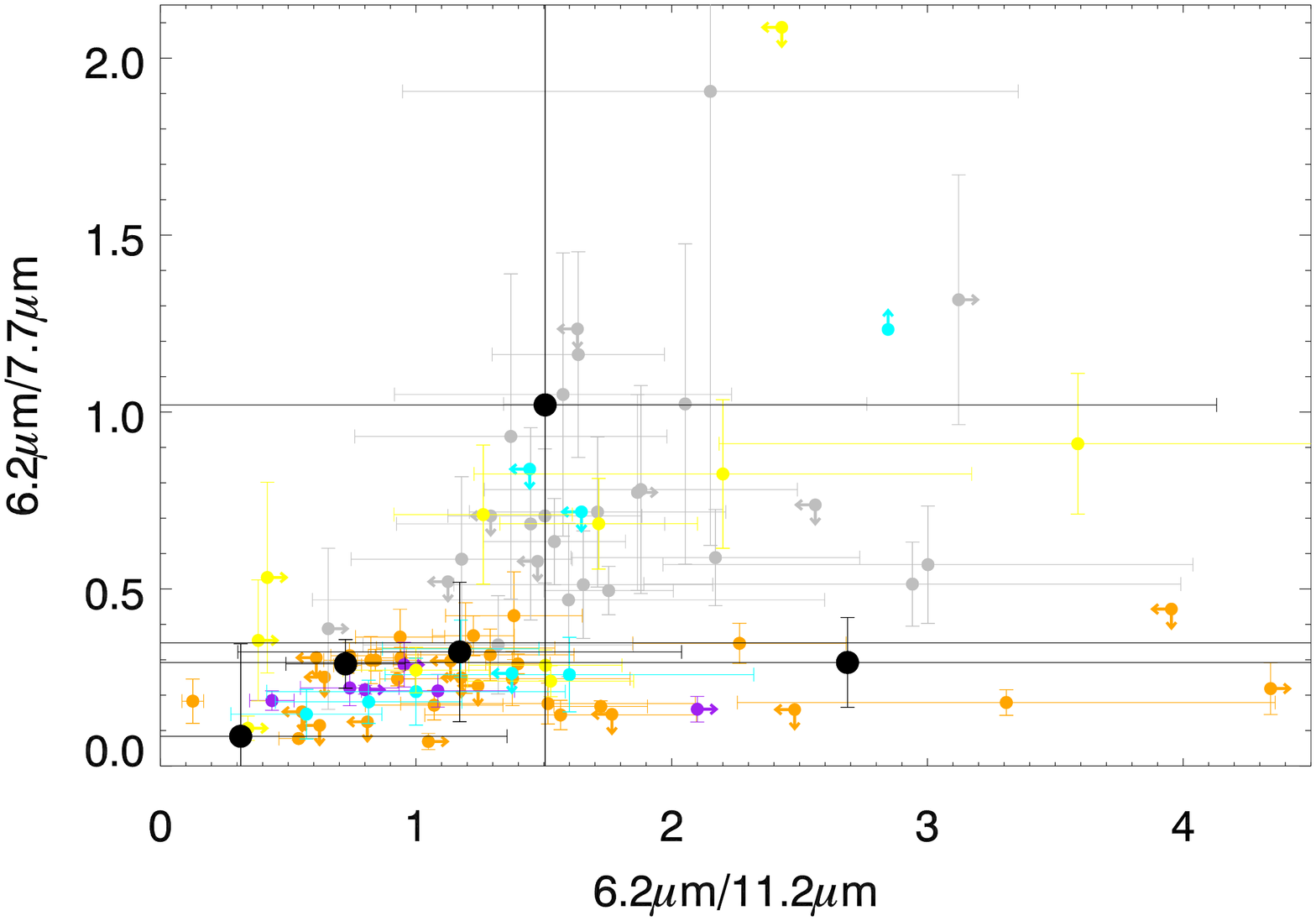}
\includegraphics[angle=90,width=85mm]{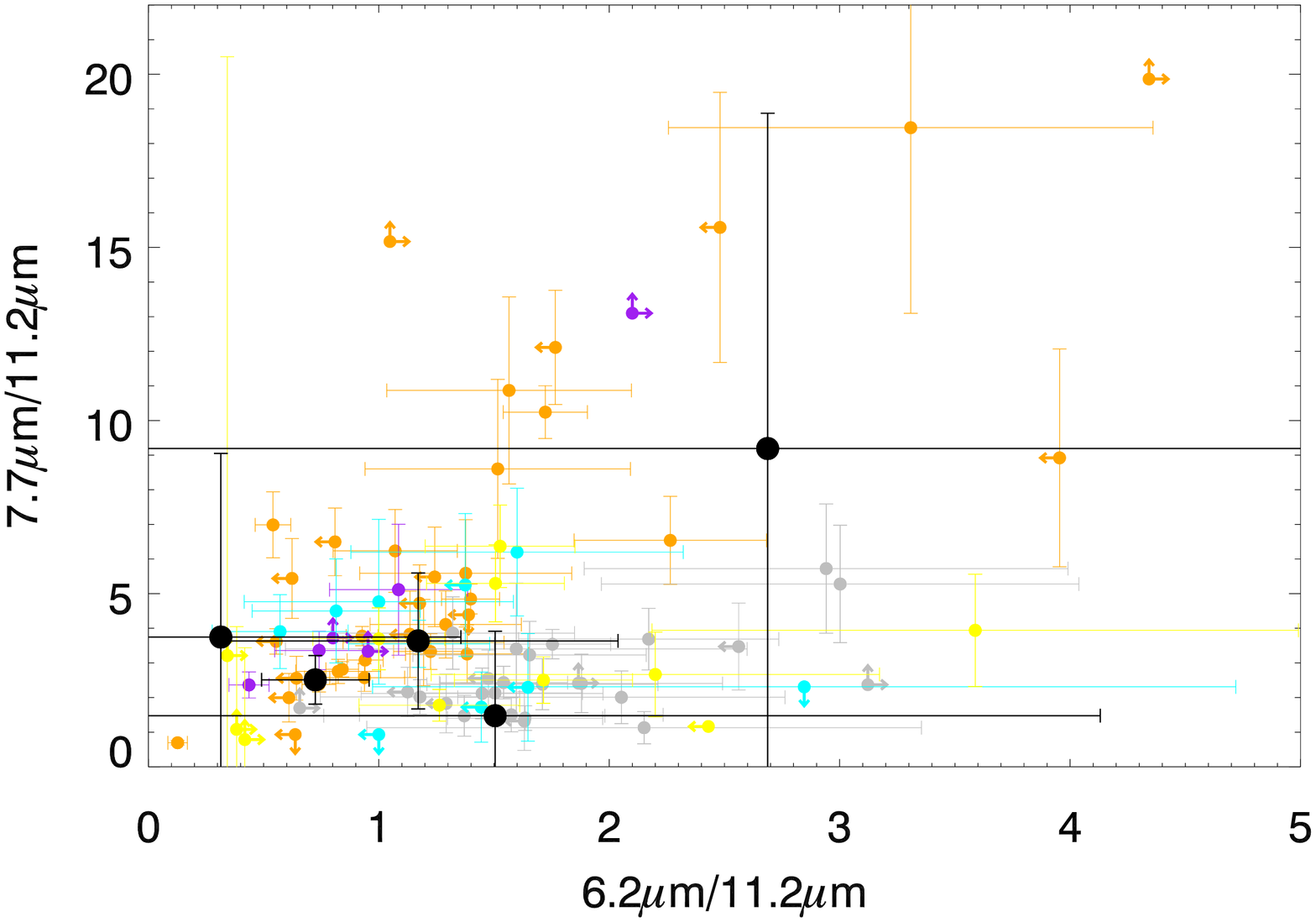}
\end{minipage}
\caption{PAH ratio diagnostic plots, divided by comparisons to (top row) low-redshift and (bottom row) high redshift samples. 
{\it Black:} FeLoBALs.
{\it Green:} low-z ULIRGs with measurable 7.7$\mu$m PAHs \citep{spo07,des07}.
{\it Red:} low-z AGN \citep{wee05}.
{\it Blue:} low-z starbursts with L$_{IR}<10^{11}$L$_{\odot}$ \citep{bdl06}.
{\it Purple:} high-z $70\mu$m selected \citep{bra08a,far09a}.
{\it Grey:} high-z  `bump' selected \citep{far08}.
{\it Orange:}  high-$24\mu$m/0.7$\mu$m selected  \citet{das09}. 
{\it Cyan:}  high-z $24\mu$m/8$\mu$m and $24\mu$m/0.7$\mu$m selected  \citet{saj07}.
{\it Yellow:}  high-z sub-mm selected \citep{pop08,men09}.
\label{pahratio}}
\end{figure}

\begin{figure}
\begin{minipage}{180mm}
\includegraphics[angle=90,width=85mm]{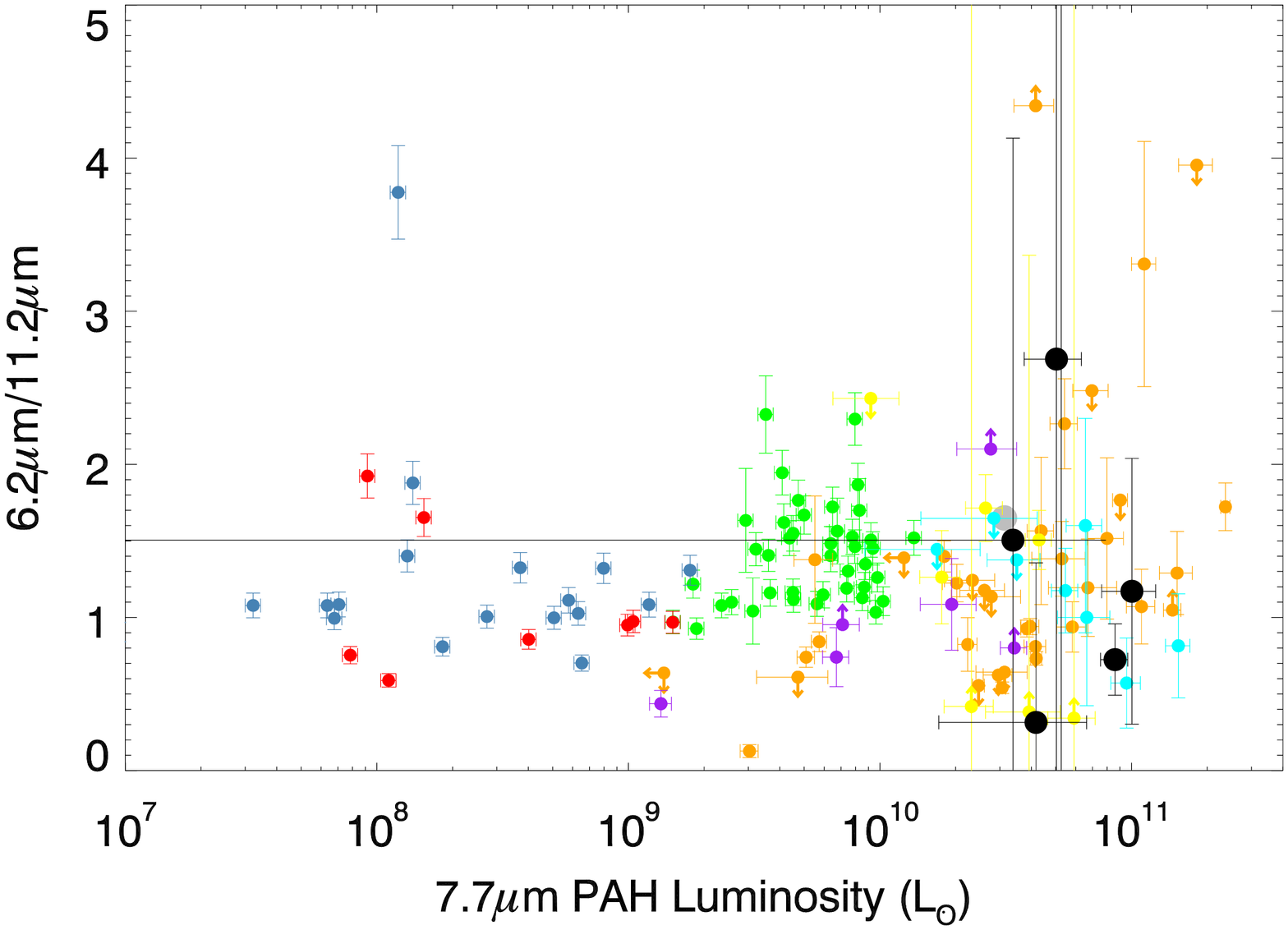}
\includegraphics[angle=90,width=85mm]{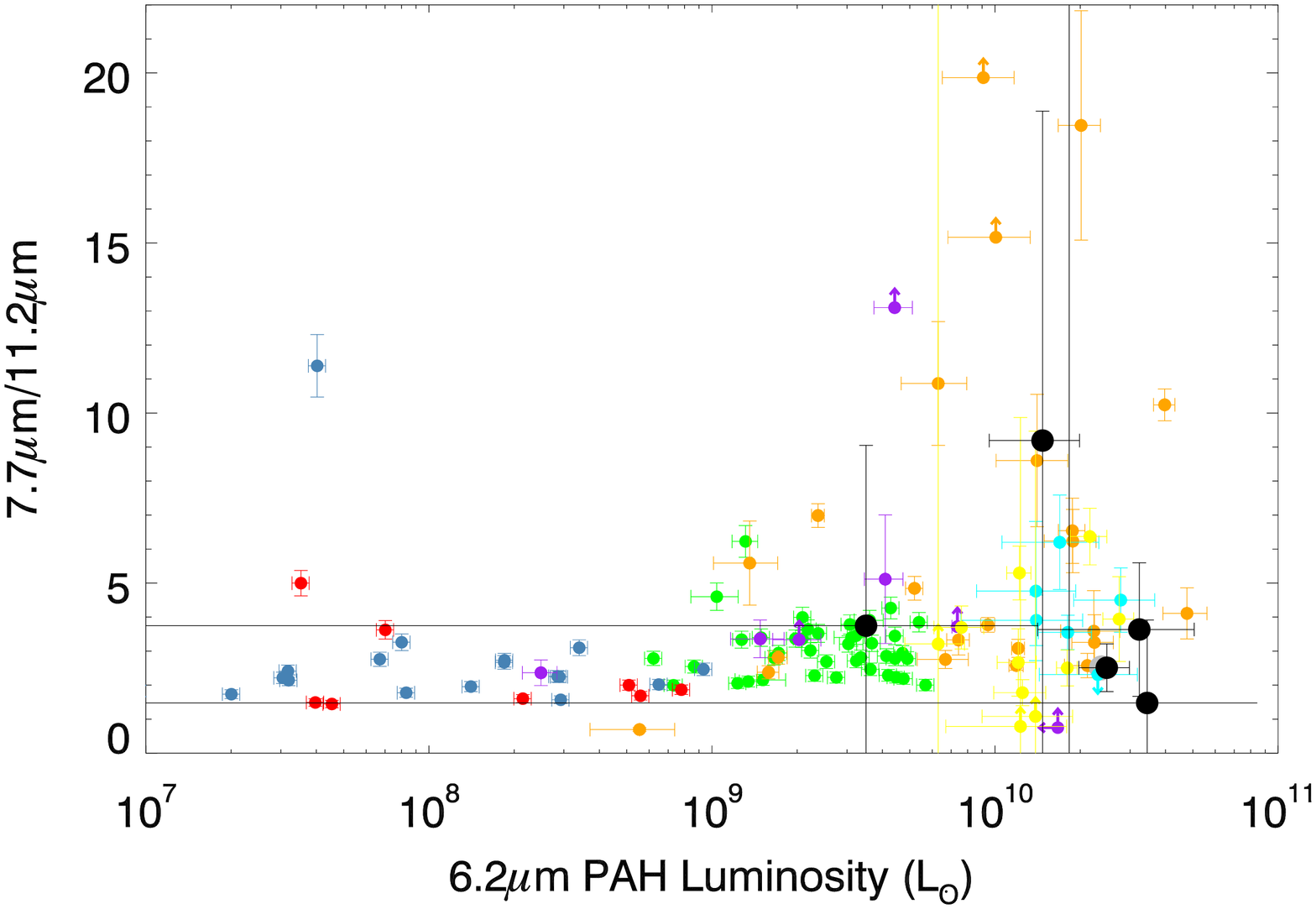}
\end{minipage}
\caption{PAH luminosity diagnostic plots. Color coding is the same as in Figure \ref{pahratio}. 
\label{pahlum}}
\end{figure}

We start with the PAH features. The PAH flux ratios of our sample, considered either as functions of each other or of PAH luminosity (Figures \ref{pahratio} and \ref{pahlum}) are slightly offset from those of low-redshift ULIRGs, but lie within the dispersion of high-redshift 70$\mu$m or 24/r selected samples. This is straighforward to understand. Local ULIRGs are selected without a bias towards AGN, and have lower IR luminosities, on average, than our sample. Conversely, the \citet{saj07} \& \citet{das09} samples are (arguably) biased towards AGN, and have comparable total IR luminosities to our objects. It is thus not surprising that the PAH flux ratios and luminosities of our sample resemble the high redshift comparison objects. It is also interesting that the two systems with the strongest PAH detections (SDSS 1214-0001 \& SDSS 1427+2709) are also the strongest 160$\mu$m detections, consistent with the idea that starbursts are associated with colder dust, but the small size of our sample means this consistency could simply be coincidence, so we do not comment on it further. 

We estimate star formation rates from the PAH features using:

\begin{equation}
SFR [M_{\odot} yr^{-1}] = 1.18\times10^{-41}L_{P} [ergs\ s^{-1}]\label{sfreq}
\end{equation}

\noindent where $L_{P}$ is the combined luminosity of the 6.2$\mu$m and 11.2$\mu$m PAH features \citep{far07b}. The values are listed in Table \ref{tabpahsils}\footnote{Estimating the star formation rates using the formula in \citet{hou07} gives comparable results}. Overall, they are comparable to those derived for other high redshift IR-luminous sources, including `bump' selected starbursts \citep{far08}, sub-mm selected starbusts \citep{pop08,men09}, and far-IR bright QSOs \citep{lut08}. The star formation rate in SDSS 1214-0001 is extraordinarily high; assuming continuous star formation then this star formation rate is capable of manufacturing $10^{11}$M$_{\odot}$ of stars in less than 100Myr, and $\lesssim 50$Myr if we assume a late stage exponentially decaying burst.

\begin{figure}
\begin{minipage}{180mm}
\includegraphics[angle=90,width=130mm]{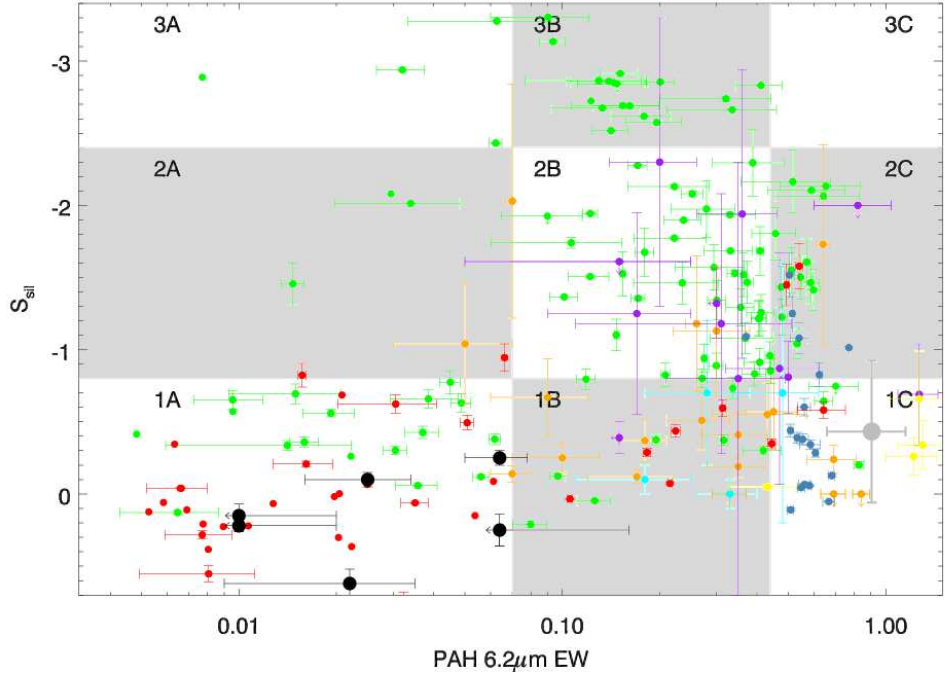}
\end{minipage}
\caption{`Fork' diagnostic diagram \citep{spo07}. The FeLoBALs are plotted in black, while the other points are color-coded as in Figure \ref{pahratio}. Plots of the composite spectra for each region are in figure 2 of \citet{spo07}. With respect to the comparison objects plotted in Figure \ref{comparisons}; Mrk 231 is class 1A, IRAS 15206+3342 is class 1B, Arp 220 is class 3B, the starburst composite is class 1C, and the remaining objects are all class 1A. The `bump' sources lie close together on this plot, so, for clarity, we plot their average as a single point.
\label{fork}}
\end{figure}

Next, we employ the `Fork' diagnostic of \citealt{spo07} (Figure \ref{fork}), which employs both the 6.2$\mu$m PAH feature and the 9.7$\mu$m silicate feature. Here our sample is not distributed in a similar way to the 70$\mu$m or 24/r selected sources, but instead lies along the lower branch of the fork. \citealt{spo07} postulate that sources move around the Fork diagram as their power source evolves, starting on the upper branch and then moving diagonally or vertically downwards as starburst/AGN activity clears the obscuration from the nuclear regions. Therefore, solely from this diagnostic, we would classify our sample as late-stage ULIRGs.

We move on to consider mid-IR continuum diagnostics. Given the limited wavelength coverage, we use the rest-frame 6$\mu$m continuum luminosity as a proxy for the bolometric AGN luminosity \citep{nar08,wat09}. Following \citealt{nar08}, we define:

\begin{equation}
R = \frac{L_{6}}{L_{IR}}
\end{equation}

\noindent where $L_{6}$ is the 6$\mu$m luminosity as measured from the IRS spectra and $L_{IR}$ is the 1-1000$\mu$m luminosity from F07. The fractional contribution of an AGN to the 6$\mu$m luminosity, $\alpha_{6}$, is then:

\begin{equation}
\alpha_{6} = \frac{1}{R_{S} - R_{A}}\left( \frac{R_{A}R_{S}}{R} - R_{A}\right)
\end{equation}

\noindent where $R_{S}$ and $R_{A}$ are the equivalents of $R$ for `pure' starbursts and AGN, with values of $(117\pm8)\times10^{-4}$ and $0.32\pm0.1$ respectively \citep{nar08}. Therefore, the fractional contribution of the AGN to the total IR luminosity, $\alpha_{bol}$, is:

\begin{equation}\label{agnbolfract}
\alpha_{bol} = \frac{\alpha_{6}}{\alpha_{6} + \left(\frac{R_{A}}{R_{S}}\right)(1-\alpha_{6})} 
\end{equation}

\noindent The resulting $\alpha_{bol}$ values are listed in Table \ref{tabsample}. For comparison, we also list the $\alpha_{bol}$ values obtained from the SEDs in F07. In four cases the values are consistent, but for two (SDSS 1556+3517 \& SDSS 2215-0045) they are not. We think it likely that the SED derived values are more reliable, as they are direct measurements from the (albeit sparsely sampled) full SEDs, while the values computed using Equation  \ref{agnbolfract} are calibrated using a wide range of sources. It is interesting that these two objects are the only two with detected silicate emission, but we do not know if this is the cause of the discrepancy, or coincidence. Considering either method though, the sources with strong PAHs have weak AGN contributions to the total IR luminosity, and the range in $\alpha_{bol}$ values for the whole sample is similar to that seen in local ULIRGs.

\begin{figure}
\begin{minipage}{180mm}
\includegraphics[angle=90,width=130mm]{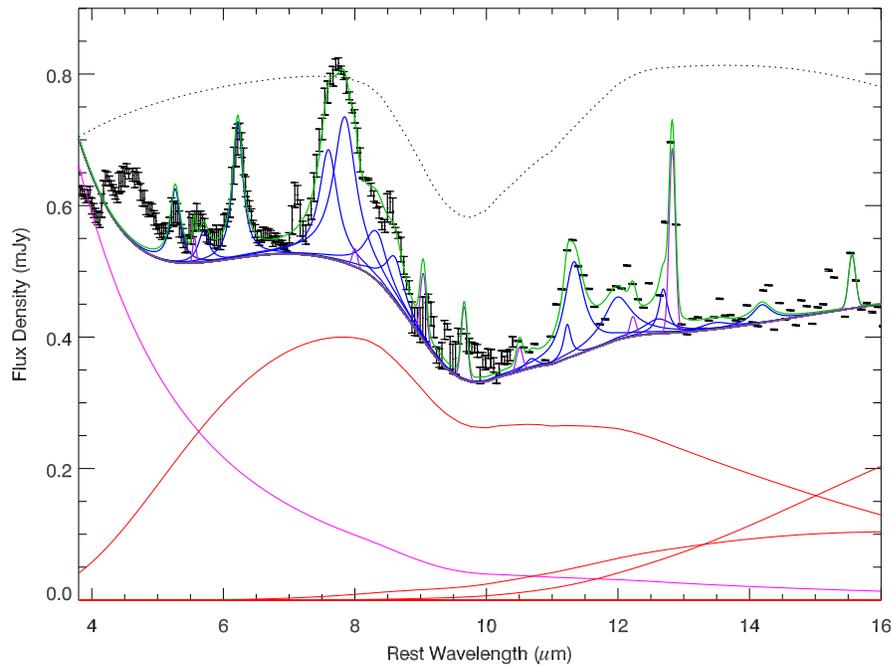}
\end{minipage}
\caption{PAHfit results for SDSS 1214-0001. Details of the model parameters can be found in \citet{jds07}.
{\it Vertical black lines:} observed data.
 {\it Green:} combined best-fit model.
 {\it Red solid lines:} thermal dust continuum components.
 {\it Magenta:} starlight continuum.
 {\it Black solid:} combined starlight+thermal dust continuum.
 {\it Violet:} fine structure+molecular features.
 {\it Blue:} PAH features.
 {\it Black dotted:} extinction curve.
\label{pahfit}}
\end{figure}

Finally, we fit the source with the most prominent PAHs, SDSS 1214-0001, with PAHFIT \citep{jds07}. This is not an attempt to reconstruct star formation parameters as PAHFIT is intended for systems where emission from stars provides at least most of the flux across the mid-IR, but it does serve as a test of how much of the mid-IR flux in SDSS 1214-0001 is attributable to star formation. The result is shown in Figure \ref{pahfit}. The fit is good, with $\chi^2_{red}=1.4$. The fit is poor at $\lesssim 6\mu$m, with a steeply rising contribution from `starlight' with decreasing wavelength that is probably an attempt to fit the AGN continuum. At $\lambda \gtrsim 6\mu$m though, we obtain an excellent fit. The PAHs are well fitted by the Drude profiles, and the continuum is predicted to mostly come from dust at 50k, with small contributions from the 135k and 300k components.

\section{Discussion \& Conclusions}
A general caveat to all of what follows is our small and heterogeneous sample. Our conclusions should thus be regarded as tentative. 

The mid-IR spectra of our sample span a wide range of shapes. We see classical QSO spectra with hot silicate dust, together with classical starburst spectra with strong PAHs. The `starburst' spectra have more prominent PAHs than those seen in any other QSO so far observed, while the hot silicate dust sources do not have a close match in any purely mid/far-IR selected sample that we are aware of. It is possible that our heterogeneous selection leads to the heterogeneity of the spectra, but if this were true then we might expect the ISO selected object to have the strongest PAHs, which is not the case. Indeed, the two spectra at the extreme ends of the range of spectral shapes are both SDSS objects. The star formation rates of our sample span levels comparable to those in the most luminous starbursts found in any survey at any wavelength, to those seen in moderately IR-luminous AGN. The spectral diagnostics mark our sample as late-stage ULIRGs with a wide range of fractional AGN contributions. 

These results, combined with their optical classification as reddened QSOs with a strong outflow, are in principle compatible with a continuum of `end-of-ULIRG' vs `peculiar QSO' contributions to the FeLoBAL QSO population. To frame the discussion, we describe three points on this continuum in detail:

\noindent {\bf 1:} FeLoBAL QSOs are a transition stage between starburst dominated ULIRGs and QSOs. As our sample span the entirety of this transition, and because FeLoBALs are observed to be rare\footnote{A further reason behind the rarity of FeLoBALs is that they they are hard to detect; FeLoBAL QSOs are red, have few or no broad emission lines and no UV excess, and are thus hard to find in standard QSO searches (see e.g. \citealt{app05}). The iron absorption features themselves are weak.}, we can set constraints on the timescale of this transition depending on how intrinsically common the Fe absorption features are. If Fe absorption is ubiquitous in such a transition, then the transition must be short in comparison to the ULIRG or QSO lifetime. Taking the ULIRG and QSO lifetimes to be $\sim$10$^{8}$ years \citep{tac08}, then the transition must take $\lesssim$10$^{7}$ years (see also \citealt{gre02}). If on the other hand the iron absorption features are intrinsically rare, then the transition can take longer. 

\noindent {\bf 2:} FeLoBALs are comprised of comparable fractions of two galaxy populations, one that is transitioning from a ULIRG to a QSO, and another that is an unusual phase that QSOs go through, unconnected to mergers. Here the heterogeneity of our sample arises from observing two unconnected populations.

\noindent {\bf 3:} FeLoBAL QSOs are entirely an `unusual QSO' class, and have nothing to do with mergers or otherwise `transitioning' systems. Instead, the iron absorption features mark QSOs with both an outflow and atypically large amounts of iron in their ISM. 

Formally, we cannot rule any of these scenarios out. The third though seems unlikely. It is hard to see how such a heterogeneous set of mid-IR spectra could arise from observing a peculiar class of QSO in which nothing more interesting than an outflow with large quantities of iron is going on, especially if we assume that the restrictive selection on rest-frame UV properties means we are viewing them over a restricted range in line of sight. Furthermore, most systems with star formation rates in the range seen in our sample are ULIRGs, which (at low redshift, at least) are nearly all mergers.

Distinguishing between the first and second scenarios is however more difficult. On one hand, the first scenario uses a single origin for which there is independent evidence from studying ULIRGs. The second scenario thus seems contrived, as it proposes two origins where one will do. On the other hand, our sample is small, and it is tempting to place undue emphasis on the properties of SDSS 1214-0001, as it is so peculiar. If this object were removed then the `unusual QSO' scenario would become more attractive. That said, the counter-argument also holds; remove (say) SDSS 1556+3517, and the `end-of-ULIRG' scenario becomes more attractive. We are mindful though that it is only the properties of SDSS 1214-0001 (and to an extent SDSS 1427+2709) that make us seriously consider the `end-of-ULIRG' scenario. 

Recent work that may shed light on this is the study of NLS1's by \citet{san10}. Their spectra show that at least some NLS1's harbor intense star formation and a weak mid-IR AGN continuum. NLS1's also have strong optical Fe emission lines, a large soft X-ray excess, and exhibit rapid, large amplitude X-ray variability. Furthermore, their black hole masses appear to be smaller than those in BLS1's of comparable luminosity, though there is controversy over whether NLS1's lie below \citep{pet00,gru04} or on \citep{bot05,kom07,dec08} the $M_{BH} - \sigma_{*}$ relation \citep{tre02}. On one level this hints at interesting links between NLS1s and FeLoBAL QSOs; for example, FeLoBAL QSOs may be analogues of NLS1s seen more pole-on than edge-on, and in which a strong outflow has formed. Exploring this idea in detail requires significantly larger samples, so we do not pursue it here. More generally however, it implies that relative orientation {\it may} also play a role in the FeLoBAL QSO phenomenon. 

We think it likely that the simplest solution that is consistent with all the observations, and the apparent rarity of FeLoBALs, is the correct one. We therefore propose that FeLoBAL QSOs as a class are rapidly evolving, youthful QSOs in which a starburst is coming to an end, and the AGN is in the last stages of burning through its surrounding dust. We also propose, with more reserve, that (1) we view FeLoBAL QSOs over a restricted line of sight range, where we see them more pole-on than edge on, and (2) the outflow has in some cases partially cleared the dust from around the AGN, leaving the IR emission from the starburst to dominate.

There are three ways to test this conclusion. First, FeLoBAL QSO hosts should be in the final stages of merging, so their host galaxies should show slight signs of morphological disturbance. Second, larger mid-IR spectroscopic surveys of FeLoBAL QSOs should show a similar range of spectral shapes to ours. Third, far-IR photometric surveys of FeLoBALs should find both a moderate enhancement of the fraction of far-IR bright sources compared to the classical QSO population, corresponding to those FeLoBAL QSOs that still harbor high star formation rates, and a correlation between far-IR luminosity and PAH equivalent width. Moreover, it would be useful to perform radiative transfer modelling of a BAL wind through a starburst to determine the conditions under which iron absorption is observed in such systems, and so provide insight into whether our conclusions are feasible. 

We close with a caveat and a speculative comment. First, there is an important parameter that we cannot address; the way in which iron absorption features in FeLoBALs signpost starburst and AGN activity. For example, if the Fe absorption only occurs when a starburst is ending {\it and} an AGN is blowing away its surrounding dust, but can occur throughout such a transition, then our first scenario is likely correct. As a contrived counter-example, if Fe absorption can only occur during the early phases of such a transition, but can occur randomly in reddened AGN in which there is no significant star formation, then in a small sample such as ours we would conclude, incorrectly, that all FeLoBAL QSOs were a ULIRG-to-QSO transition where the Fe absorption is present throughout. Resolving this issue is beyond the scope of this paper, so we note it as a caveat to our conclusions. 

Second, \citet{hall02} note that two of our sample, SDSS 1214-0001 and SDSS 2215-0045 have [FeIII]/[FeII] ratios greater than unity, implying the BALs arise in unusually dense, hot gas. The two ratios however are different; SDSS 2215-0045 has a [FeIII]/[FeII] ratio much greater than unity, while SDSS 1214-0001 has an [FeIII]/[FeII] ratio only slightly greater than unity. While conclusions drawn on only two objects are not trustworthy, this difference in ratio is consistent with our proposed evolutionary sequence. SDSS 1214-0001 has strong PAHs and silicates in absorption, while SDSS 2215-0045 has weak PAHs and silicates in emission, implying that SDSS 1214-0001 is at an earlier evolutionary phase than SDSS 2215-0045. Moreover, the enhanced [FeIII]/[FeII] ratio in SDSS 2215-0045 compared to SDSS 1214-0001 could be interpreted as the BAL outflow in SDSS 2215-0045 being more developed\footnote{This is consistent with the `youth' hypothesis that \citet{hall02} give for SDSS 2215-0045}. Further support for this idea would be an [FeIII]/[FeII] ratio greater than unity in SDSS 1427+2709, but the only published spectrum we are aware of is in \citet{bec00}, in which the FeIII lines lie in a noisy region at the extreme blue end of the bandpass. They do not however appear to be significantly stronger than the FeII lines, so we simply note this as an interesting avenue for future work.

\acknowledgments
We thank the referee for a helpful report. This work is based on observations made with the Spitzer Space Telescope, which is operated by the Jet Propulsion Laboratory, California Institute of Technology under a contract with NASA. Support for this work was provided by NASA. This research has made extensive use of the NASA/IPAC Extragalactic Database (NED) which is operated by the Jet Propulsion Laboratory, California Institute of Technology, under contract with NASA. DF thanks STFC for support via an Advanced Fellowship. NC gratefully acknowledges support from a Cottrell College Science Award from the Research Corporation and from an NSF RUI grant.

\begin{deluxetable}{lcccccccccc}
\tabletypesize{\scriptsize}
\tablecolumns{11}
\tablewidth{0pc}
\tablecaption{FeLoBAL QSO Sample \label{tabsample}}
\tablehead{
\colhead{Object}&\colhead{RA (2000)}&\colhead{Dec}&\colhead{$z_{sys}$\tablenotemark{a}}&\colhead{$z_{abs}$\tablenotemark{b}}&\colhead{$f_{24}$}&\colhead{$f_{70}$}&\colhead{$f_{160}$} &\colhead{$\alpha_{bol}|_{6\mu m}$\tablenotemark{d}}&\colhead{$\alpha_{bol}|_{sed}$\tablenotemark{e}}&\colhead{$L_{IR}$}
}
\startdata
(1) ISO  J005645.1-273816           & 00 56 45.2  & -27 38 15.6   & 1.78                                  & 1.75 & 1.6     & $<$7.0 & $<$50  & 0.09 - 0.40  & $>0.26$ &  12.5-13.0  \\
(2) SDSS J115436.60+030006.3  & 11 54 36.6 & 03 00 06.4     & 1.46\tablenotemark{c}& 1.36 & 7.6     &   17.9    & $<$50  & 0.35 - 0.60  & $>0.40$ &  12.9-13.1  \\
(3) SDSS J121441.42-000137.8   & 12 14 41.4 & -00 01 37.9    & 1.05                                  & 0.99 & 4.7     &   38.3     &    78.9  & 0.07 - 0.15  & $<0.37$ & 12.7-12.9   \\
(4) SDSS J142703.62+270940.3  & 14 27 03.6 & 27 09 40.3     & 1.17                                 &  ?      & 4.8     &   32.6    &    68.1  & 0.09 - 0.17  & $<0.61$ &  12.8-13.0  \\
(5) SDSS J155633.78+351757.3   & 15 56 33.8 & 35 17 58.0      & 1.50                                 & 1.48  & 13.9   &   24.7    & $<$50  & 0.40 - 0.70  & $>0.75$ &  13.1-13.3 \\
(6) SDSS J221511.93-004549.9    & 22 15 11.9 & -00 45 49.9    & 1.48                                 &  1.36 & 10.4   &   27.2    & $<$50  & 0.09 - 0.30  & $>0.36$ &  13.0-13.4  \\
\enddata
\tablecomments{All flux densities are quoted in mJy. Errors are typically 10\% at 24$\mu$m, 20\% at 70$\mu$m, and 25\% at 160$\mu$m. Luminosities are the logarithm of the rest-frame 1-1000$\mu$m luminosity, in units of solar luminosities (3.826$\times10^{26}$ Watts), taken from F07. Limits and ranges are 3$\sigma$.}
\tablenotetext{a}{Systemic redshift from narrow optical emission lines}
\tablenotetext{b}{Redshift of peak absorption from the broad UV absorption lines}
\tablenotetext{c}{\citet{hall02}}
\tablenotetext{d}{Fractional AGN luminosity, computed using the prescription in \S\ref{specfeat} \citep{nar08}.}
\tablenotetext{e}{Fractional AGN luminosity, derived from the SED fits in F07. }
\end{deluxetable}

\begin{deluxetable}{lcccccccc}
\tabletypesize{\scriptsize}
\tablecolumns{9}
\tablewidth{0pc}
\tablecaption{Spectral Measurements\label{tabpahsils}}
\tablehead{
\colhead{Object}&
\multicolumn{2}{c}{PAH 6.2$\mu$m}&
\multicolumn{2}{c}{PAH 7.7$\mu$m}&
\multicolumn{2}{c}{PAH 11.2$\mu$m}&
\colhead{$S_{sil}$}&
\colhead{SFR\tablenotemark{a}}\\
\colhead{  }&
\colhead{Flux}&
\colhead{EW}&
\colhead{Flux}&
\colhead{EW}&
\colhead{Flux}&
\colhead{EW}&
\colhead{}&
\colhead{M$_{\odot}$ yr$^{-1}$}
}
\startdata  
 (1) ISO 0056-2738       & 6.1$\pm$8.9  & 0.06$\pm$0.09 &     6.0$\pm$8.0  &  0.08$\pm$0.12 &   4.1$\pm$3.9  & 0.08$\pm$0.08 & 0.25$\pm$0.11 & $2600\pm2500$  \\
 (2) SDSS 1154+0300  & 5.2$\pm$5.7   & 0.01$\pm$0.01 &   15.1$\pm$9.2  &  0.04$\pm$0.02 &   0.3$\pm$4.6 & 0.01$\pm$0.02 & 0.15$\pm$0.08  & $900\pm1100$    \\
 (3) SDSS 1214-0001   & 16.3$\pm$3.3 & 0.06$\pm$0.01 &   56.5$\pm$7.0  &  0.25$\pm$0.05 & 22.5$\pm$5.6 & 0.24$\pm$0.06 & -0.25$\pm$0.05  & $2700\pm500$   \\
 (4) SDSS 1427+2709 & 7.3$\pm$2.6   & 0.03$\pm$0.01 &   25.0$\pm$6.4  &  0.10$\pm$0.03 &   2.4$\pm$2.1 & 0.02$\pm$0.01 & -0.10$\pm$0.05 & $900\pm350$      \\
 (5) SDSS 1556+3517  & 1.0$\pm$3.0   & 0.01$\pm$0.01 &   11.2$\pm$6.6  &  0.02$\pm$0.01 &   3.0$\pm$3.9 & 0.01$\pm$0.02 & 0.22$\pm$0.04  & $700\pm800$      \\
 (6) SDSS 2215-0045   & 10.5$\pm$5.3  & 0.03$\pm$0.01 &   28.1$\pm$6.8  & 0.09$\pm$0.02 &   7.7$\pm$3.7 & 0.02$\pm$0.01 & 0.62$\pm$0.10   & $2700\pm1000$  \\
\enddata  
\tablecomments{PAH fluxes are in units of $10^{-22}$W cm$^{-2}$ and equivalent widths in $\mu$m.}
\tablenotetext{a}{Star formation rate, determined from Equation \ref{sfreq}. Error is solely that derived from the uncertainty in the fluxes. }
\end{deluxetable}

\end{document}